\documentclass[12pt,preprint]{aastex}

\slugcomment{To appear in ApJ Letters}

\shorttitle{Determation of the inclination of HR8799}
\shortauthors{Wright et al.}

\begin{document}

\title{Determination of the inclination of the multi-planet hosting star HR8799 using asteroseismology
\footnote{Based on observations obtained at the Observatoire de Haute-Provence which is operated by the Institut National des Sciences de l'Univers of the Centre National de la Recherche Scientifique of France.}
}

\author{D.~J. Wright}
\affil{Koninklijke Sterrenwacht van Belgi\"e, Ringlaan 3,\\
 B-1180 Brussel, Belgium}
\email{Duncan.Wright@oma.be}

\author{A.-N. Chen\'e}
\affil{Canadian Gemini Office, HIA/NRC of Canada, \\
5071, West Saanich Road, Victoria (BC), V9E 2E7, Canada}
\affil{Departamento de Astronom\'ia, Casilla 160-C, Universidad de Concepci\'on, Chile}
\affil{Departamento de F\'isica y Astronom\'ia, Facultad de Ciencias,\\ 
Universidad de Valpara\'iso, Av. Gran Breta\~na 1111, Playa Ancha,\\
Casilla 5030, Valpara\'iso, Chile}
\email{achene@astro-udec.cl}

\author{P. De Cat}
\affil{Koninklijke Sterrenwacht van Belgi\"e, Ringlaan 3,\\
 B-1180 Brussel, Belgium}
\email{peter@oma.be}

\author{C. Marois}
\affil{National Research Council Canada, Herzberg Institute of Astrophysics,\\
5071 West Saanich Road, Victoria, BC V9E 2E7, Canada}
\email{Christian.Marois@nrc-cnrc.gc.ca}


\author{P. Mathias}
\affil{Laboratoire d'Astrophysique de Toulouse-Tarbes, Universit\'e de Toulouse, CNRS, 57 avenue d'Azereix, F-65000 Tarbes, France}
\email{pmathias@ast.obs-mip.fr}


\author{B. Macintosh}
\affil{Lawrence Livermore National Laboratory, \\
7000 East Avenue, Livermore, CA 94550, USA}
\email{macintosh1@llnl.gov}

\author{J. Isaacs}
\affil{Lawrence Livermore National Laboratory, \\
7000 East Avenue, Livermore, CA 94550, USA}
\affil{Department of Physics, University of Wisconsin, \\
Madison, WI, 53706, USA}
\email{jaisaacs@wisc.edu}

\and


\author{H. Lehmann and M. Hartmann}
\affil{Th\"uringer Landessternwarte Tautenburg, Sternwarte 5, D-07778 Tautenburg, Germany}
\email{artie@tls-tautenburg.de; michael@tls-tautenburg.de}

\begin{abstract}
Direct imaging of the HR8799 system was a major achievement in the study of exoplanets. HR8799 is a $\gamma$\,Doradus variable and asteroseismology can provide an independent constraint on the inclination. Using 650 high signal-to-noise, high resolution, full visual wavelength spectroscopic observations obtained over two weeks at Observatoire de Haute Provence (OHP) with the SOPHIE spectrograph we find that the main frequency in the radial velocity data is 1.9875 d$^{-1}$. This frequency corresponds to the main frequency as found in previous photometric observations. Using the FAMIAS software to identify the pulsation modes, we find this frequency is a prograde $\ell$=1 sectoral mode and obtain the constraint that inclination $i\gtrsim$40$^{\circ}$.
\end{abstract}


\section{Introduction}
The imaging discovery of the three \citep{Ma08}, and now four \citep{Mr10} planets around HR~8799 is a significant achievement in the search for and study of planets orbiting other stars. For the first time, the thermal emission of planets in orbit around another star has been unambiguously detected.

The dynamical evolution of a planetary system is complex. From the basic planet formation assumption that planets form by the core accretion or disk instability scenario in a disk along the star's equatorial plane, systems can suffer drastic changes; planet-planet perturbations, interactions with a disk or stellar encounters can change a planet's orbital inclination, its semi-major axis, its orbital eccentricity or even eject it \citep{Ra10}. In the case of HR~8799, it is not impossible that a close encounter occurred prior to planet formation, since its relatively high galactic velocity compared to the Columba association and its far distance away from the other association members \citep{Hi10} are suggesting that it may have been kicked out and is probably moving quickly away from its birth place. Such an encounter could have tilted a disk relative to the star's equatorial plane and induce perturbations that may have led to planet formation where planets would have a non-negligible orbital inclination relative to the star. Radial velocity searches have confirmed such chaotic behaviors by detecting systems where planets are orbiting well away from the star's equatorial plane (\citealt{Tr10} and references therein), although for close-in extrasolar planets this misalignment could also be caused by the Kozai mechanism \citep[e.g.][]{Wu07,Fa07,Wi09}. For the wide HR8799 planets the Kozai mechanism is not operational (even if the system had a stellar companion in the past); finding a misalignment between the star and the planet's orbital plane would be a sign of a significant dynamical interaction in the system past.

\citet{Ma08} have suggested that the HR~8799 planets are in a similar orbital plane with a low inclination and have mostly circular orbits. This is because the detected orbital motions are close to their expected face-on circular orbit values, the orbital motion is mainly in azimuth and the star is known to be a slow rotator (thus it would be viewed mainly pole-on). Dynamical analyses \citep[e.g. ][]{Re09,Fa10,Mo10,Ma10} have confirmed that the planets are mostly in the same plane with small eccentricities, although such fits are still very uncertain due to the limited amount of orbital coverage available. In addition, the planets also all orbit in the same counter-clockwise orientation, further supporting the idea that they formed in a disk, similar to the Solar system planets. Assuming a circular orbit for b, \citet{La09} found $i\sim$13$^\circ$--23$^\circ$; while attempting a coherent analysis of various portions of observational data on known components of the system, \citet{Re09} concluded that $i$ should range between 20$^\circ$ and 30$^\circ$. Also,  Spitzer observations of HR~8799's complex debris disk suggest that any inclination angle larger than $\sim$25$^\circ$ should be excluded \citep{Su09}. Using a statistical distribution of star's rotation speed \citep{Ro07}, HR~8799 with its 37.5$\pm$2 km\,s$^{-1}$ V\,sin\,$i$ \citep{Ka98} would be consistent with an inclination of $\sim 23.5^\circ$ if it is an A5 star or $\sim 18.5^\circ$ if it is an F0 star (HR~8799 spectral classification is uncertain mainly due to its low metallicity that is affecting it's broad band colors). Such a determination is of course statistical and a direct star's inclination determination is required for a meaningful comparison with the estimated planet orbital plane inclination.

HR~8799 is an intrinsic photometric and spectroscopic variable \citep{Ro95,Ma04}. It has also been confirmed as a $\gamma$\,Doradus (Dor) variable \citep{Ze99}. The $\gamma$\,Dor stars are late A to early F stars whose pulsations are driven by a flux-blocking mechanism at the base of their convective envelope \citep[e.g.][]{Du04}. The $\gamma$\,Dor nature of HR~8799 offers a unique opportunity to estimate its inclination via an asteroseismic analysis of the observed $g$-modes. In a previous asteroseismic analysis using photometric frequencies \citet{My10} has shown that an age determination for the system, which allows to discriminate between planets and brown dwarfs, is a difficult task with the current information and discussed the importance of an inclination determination since the equatorial rotational velocity can be used in constraining the age of the system.
In this letter, we present a spectroscopic asteroseismic analysis and also obtain limits for HR~8799's stellar inclination. The data have been acquired from an extensive multi-site ground-based high-resolution spectroscopy campaign. Section~2 describes that data used in this letter. Section~3 discusses the pulsation mode identification and the determination of the stellar inclination and Section~4 describes the conclusions.

\section{Observations}
In this letter we examine data collected from Observatoire de Haute Provence (OHP) using the SOPHIE spectrograph. We have 650 SOPHIE observations taken from 28/09 - 12/10/2009. They cover a wavelength range between 3875--6940\AA~with a spectral resolution of R$\sim$75\,000. These data are the largest single-site subset of more than 2000 observations taken during an intensive spectroscopic multi-site campaign devoted to this star. Only the OHP spectra will be examined in this publication as study of the complete combined data set of observations will be left until the detailed publication.

The data were treated using the local reduction packages written specifically for the SOPHIE spectrograph data. 
This produced reduced, merged and automatically continuum normalized 1-D spectra which were further manually normalized using a synthetic spectrum for guidance.

Instead of using single spectral lines, we examined the cross-correlation profile (CCP). This profile is obtained using the normalised 1-D spectra cross-correlated with a line mask. The line mask was produced with the SYNSPEC software \citep{Hu95} using a T$_{eff}$=7500 K, $\log(g)=4.0$ and Z=-1.0. All the telluric and H/He affected areas were removed from the spectral line for which we estimated the equivalent widths. These equivalent widths were used for the weightings given to the delta-functions in the cross-correlation. 
An analysis of the variation of several isolated spectral lines gave similar results to those outlined for the CCPs below. Since the signal-to-noise ratio of the individual spectral lines is much lower than that of the CCPs only the CCPs are discussed here. 

The resulting line variation was examined using the first moment of the line profile which is similar to the radial velocity of the line (shown in Figure 1: left). A Fourier analysis of the first moment gives two strong frequencies 1.9875\,d$^{-1}$ and 1.7475\,d$^{-1}$ (shown in Figure 1: right). These frequencies by themselves produce a good fit to the observed radial velocity variations. They closely correspond with those found in photometry by both \citet{Ze99} and \citet{Cu06}.

\section{Pulsation mode identification}

It is possible to identify further frequencies present in the data but upon examination it is observed that extraction of any combination of the other frequencies present does not alter the results for the 1.98\,d$^{-1}$ frequency. The focus for this letter is on the inclination restrictions possible from the 1.98\,d$^{-1}$ frequency, hence we will continue without examining any further frequency information, and leave that for the detailed publication based on the whole dataset of the campaign.

\begin{figure}[!ht]
\centering
\includegraphics[scale=0.3]{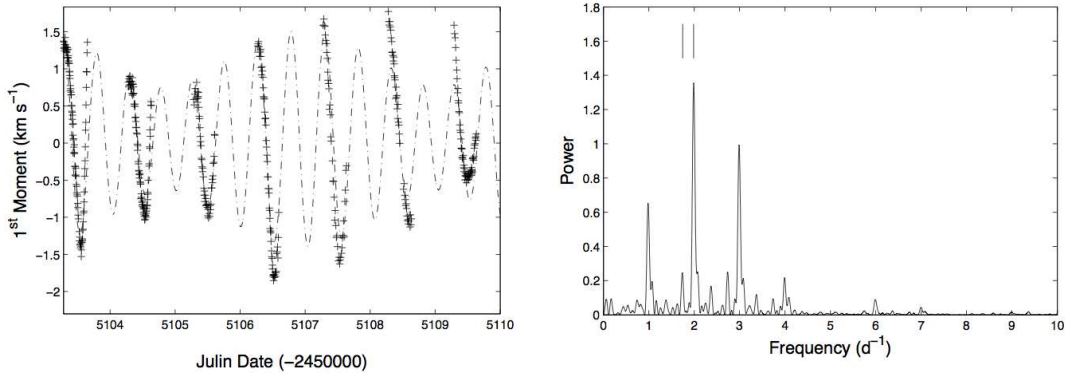}
\caption{OHP Radial velocity data and 1.9875\,d$^{-1}$ and 1.7475\,d$^{-1}$ fit for the first half of the OHP data (left). Frequency analysis of the radial velocity of the OHP data (right) with lines marking the frequencies determined.}
\end{figure}

To determine the degree $\ell$ and azimuthal order $m$ of the pulsation mode using the FAMIAS software\footnote{The software package FAMIAS developed in the framework of the FP6 European Coordination Action HELAS (http://www.helas-eu.org/)} \citep{Zi08} we extract the Fourier parameters, i.e. the zero point, the amplitude and the phase distributions across the CCPs, for the two frequencies. The Fourier parameters of 1.9875\,d$^{-1}$  are shown as a solid line in the panels of Figure 2. The Fourier parameters of 1.7475\,d$^{-1}$ were not sufficiently well matched by the FAMIAS software to consider this pulsation mode identified or constrained. This is probably a result of other pulsation frequencies present in the data that have not been detected or removed, and that have sufficient amplitude to affect the parameters of the 1.7475\,d$^{-1}$ mode.

The FAMIAS software uses a first order Coriolis force approximation, limiting the pulsation models that can be used to fit the Fourier parameters to those respecting $\nu \leq$ 1, where $\nu$ is the so-called ``spin-parameter''. This parameter is defined as $\nu$ = 2 f$_{\Omega}$/f$_{corot}$, where f$_{\Omega}$ is the rotation frequency and f$_{corot}$ is the pulsation frequency in the corotating frame. f$_{corot}$ is connected to the observed pulsation frequency f$_{obs}$ through: f$_{corot}$ = f$_{obs}$ - $m$f$_{\Omega}$ and, hence, changes for different values of $m$ of the model being fitted. As for f$_{\Omega}$, its changes depend on the inclination used by the model. On the other hand, one could wonder how significant is the effect of stellar rotation when $\nu >$ 1 and if the fit of the Fourier parameters could still be trusted passed this limit. \citet{To03} examined the effects of the increasing Coriolis force on high order ($n$) low degree ($\ell$) gravity-modes (see Figure 1 of \citealp{To03}) and found that prograde modes ($m>$0, this paper; $m<$0 in \citealp{To03}) suffer reduced changes from increasing Coriolis force. Using this study we can estimate more appropriate $\nu$ limitations for the FAMIAS models of the 1.9875\,d$^{-1}$  frequency. For the values of $m$=-3:3, limits of $\nu\leq$8 for $m$=3, $\nu\leq$5 for $m$=2, $\nu\leq$2 for $m$=1 and $\nu\leq$1 for $m\leq$0 are reasonable. Using these $\nu$ limitations and the estimated stellar parameters R$_\ast$=1.5\,R$_\odot$ \citep{Gr99}, V\,sin\,$i$=39.5 km\,s$^{-1}$ \citep[i.e. a value within the uncertainties given by this study and by][]{Ka98} we can place restrictions on the inclination for which the FAMIAS models are valid for each value of $m$, they are $i$= 60$^{\circ}$, 37$^{\circ}$, 32$^{\circ}$, 32$^{\circ}$, 15$^{\circ}$, 0$^{\circ}$ and 0$^{\circ}$ for $m$= 3, 2, 1, 0, -1, -2 and -3, respectively. 

These model limits imply that we cannot test all the inclinations between 0$^\circ$ and 90$^\circ$. However, one has to consider that an inclination lower than 5$^{\circ}$ is physically impossible, since it would imply an equatorial rotation velocity higher than the break-up velocity when using an estimated mass of M$_\ast$=1.5\,M$_\odot$ \citep{Gr99}. Moreover, models with $m$=1, 2 and 3 are limited to $i>$16$^{\circ}$, 33$^{\circ}$ and 54$^{\circ}$ respectively to be physically possible (f$_{corot}>$ 0). A summary of the limitations on the inclination of the pulsational axis is given in Table 1. From this table it can be seen which are the physically possible values of $m$ and inclination that we are unable to test due to the FAMIAS model restrictions. However, these ``holes'' can be mitigated by the fact that at lower inclinations than those imposed by the model limits, the $\nu$ value, and hence the Coriolis force, becomes large enough that the pulsations surface deformations begin to be limited to an equatorial waveguide \citep{To03}. In these cases just the regions about the equator are varying, but we would be viewing the star from low inclinations. This means that only a small area of the visible surface would be experiencing pulsation and the radial velocity amplitude would be expected to be low when compared with observed radial velocity amplitudes in other $\gamma$\,Dor stars. To give an example of this, if $i$=20$^{\circ}$ and the mode was $m$=0 then approximately one quarter of the full surface variability is visible since more than half the pulsation amplitude is constrained to within 35$^{\circ}$ of the equator. In contrast, the 1.98\,d$^{-1}$ frequency in question has an amplitude of 1.09 km\,s$^{-1}$ which is comparable to the radial velocity amplitudes of the strongest mode in other $\gamma$\,Dor stars e.g. 1.3 km\,s$^{-1}$ in $\gamma$\,Doradus \citep{Bl96}, 0.35 km\,s$^{-1}$ in HD49434 \citep{Ut08}, 1.45 km\,s$^{-1}$ in HD189631 and 0.49 km\,s$^{-1}$ in HD40745 \citep{Ms10}. Therefore, it is unlikely that we are dealing with a strong Coriolis force and hence a high $\nu$ value for a given $m$ value. Because of this the inclination ``holes'' mentioned previously, which are associated with high $\nu$ values, can be considered as unlikely solutions.

\vspace{10pt}
\begin{table*}[!ht]
\caption{Table of constraints on the inclination from both the model limitations and physical constraints}
\centering
\begin{tabular}{lccccccc}
$m$ value				&3	&2	&1	&0	&-1	&-2	&-3	\\
\hline 
model $i$ limits ($^{\circ}$)		&$>$60	&$>$37	&$>$32	&$>$32	&$>$15	&$>$0	&$>$0	\\
physical $i$ limits ($^{\circ}$)	&$>$54	&$>$33	&$>$16	&$>$5	&$>$5	&$>$5	&$>$5	\\

\end{tabular}
\end{table*}
\vspace{10pt}

All $\ell$ and $m$ combinations from $\ell$=0 to $\ell$=3 were tested keeping in mind the restrictions from Table 1. Higher values of $\ell$ were not considered because the 1.9875 d$^{-1}$ is a strong photometric frequency found in the data of \cite{Ze99} and modes with $\ell$\,$>$ 3 are hard to detect in such ground-based photometric studies due to geometric cancellation effects. The input parameters to FAMIAS were permitted to vary as shown in Table 2. The two best fitting modes with different $\ell$ and $m$ values obtained with FAMIAS are shown in Figure 2. The dashed-dotted line corresponds to an $\ell$=1 $m$=1 mode with $\chi^2_{red}$=30.0 and the dotted line to an $\ell$=2 $m$=-2 mode with $\chi^2_{red}$=43.9. This figure confirms that the best fitting mode is the $\ell$=1 sectoral mode as the $\ell$=2 $m$=-2 solution does not match the shape of the amplitude distribution well. 
The extremely small uncertainties in the zero point distribution cause any deviation from a perfect fit to rapidly increase the $\chi^2_{red}$ value. Hence, the value of $\chi^2_{red}$=30.0 should be considered as a very good fit. On examination of the amplitude and phase distributions across the profile for all of the modes tested, only the $\ell$=1 $m$=1 mode demonstrates a good match in shape for both and is considered conclusively the best solution for this pulsation frequency. The values for the input parameters to FAMIAS resulting from this mode are given in the last column of Table\,2. The errors are estimated based on an examination of the $\chi^2_{red}$ results.

\begin{table*}[!ht]
\caption{Table of parameters for genetic algorithm search. Both the allowed range and best fit results, an $\ell$=1 $m$=1 mode, are shown. Uncertainties are estimated based on $\chi^2_{red}$ results}
\centering
\begin{tabular}{lcc}
Parameters				&Allowed ranges		&Best fit	\\
					&min : max		&		\\
\hline 
M$_\ast$ (M$_\odot$)			&1.2 : 1.8		&1.5$\pm$0.3		\\
R$_\ast$ (R$_\odot$)			&1.2 : 1.8		&1.5$\pm$0.3		\\
Inclination ($^{\circ}$)		&5 : 90			&65$\pm$25		\\
V\,sin\,$i$ (km\,s$^{-1}$)		&36 : 46		&39.8$\pm$0.4		\\
intrinsic width (km\,s$^{-1}$)		&5 : 15			&8.3$\pm$0.5		\\
pulsation amplitude (km\,s$^{-1}$)	&0.1 : 10		&0.659$\pm$0.05		\\
pulsation phase	(2$\pi$)		&0 : 1			&0.95$\pm$0.03		\\
%
\end{tabular} 		
\end{table*}

\begin{figure}[!ht]
\centering
\includegraphics[scale=0.6]{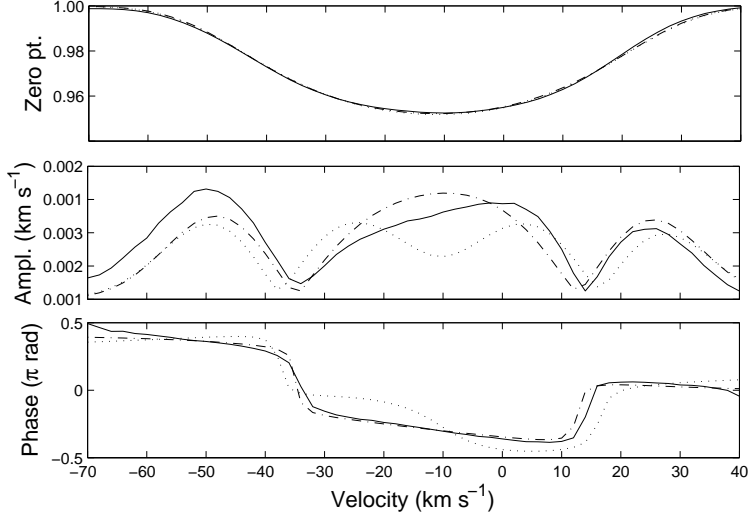}
\caption{Zero point profile (top), amplitude distribution (middle) and phase distribution (bottom) for the frequency 1.98\,d$^{-1}$ is the solid line. The best fit $\ell$=1 $m$=1 mode is the dashed-dotted line with a $\chi^2_{red}$ of 30.0. The next best fitting mode of those tested is the $\ell$=2 $m$=-2 mode shown as a dotted line with $\chi^2_{red}$=43.9.}
\end{figure} 

\begin{figure}[!ht]
\centering
\includegraphics[scale=0.6]{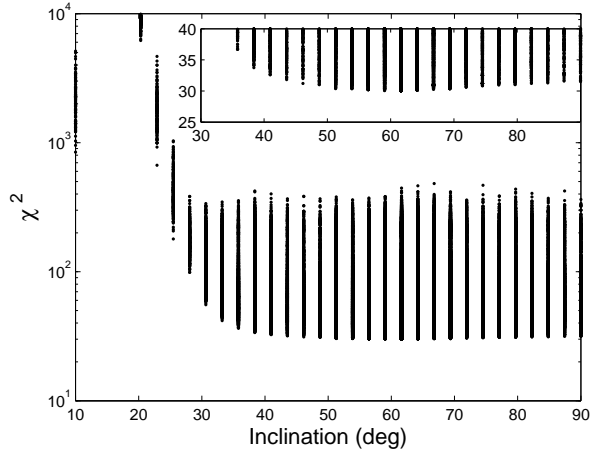}
\caption{The inclination dependence of the genetic algorithm search for the best fitting $\ell$=1 $m$=1 mode. The inset figure is a zoom in on the region of low $\chi^2_{red}$. For inclinations i$\gtrsim$40$^\circ$ the differences in $\chi^2_{red}$ is small.}
\end{figure} 


%

Figure 3 shows the inclination dependence of the genetic algorithm search for the best fitting $\ell$=1 $m$=1 mode. Based on Table 1 we can confidently use the FAMIAS results for $i\geq$32$^{\circ}$. The best fit inclination has $i$=65$^{\circ}$. For $i$= 35$^{\circ}$, 40$^{\circ}$ and 90$^{\circ}$, we find $\chi^2_{red}$= 37.2, 32.9 and 31.07, respectively. It is obvious from Figure 3 that  the $\chi^2_{red}$ does not change much in the range $i$= 40$^{\circ}$ to 90$^{\circ}$ but it rises quickly at lower inclinations. From this we obtain our result $i\gtrsim$40$^{\circ}$.

The FAMIAS software assumes that pulsation axis is aligned with the rotational axis, but this is not necessarily the case in all stars. However, such misalignment is unlikely in the case of HR~8799 because, in that case, some modulation of the pulsation properties by the rotational period would have been observed. In addition, it would have led to more complex amplitude and phase across the profile, and the fits that we have achieved would not have been as good.

It is worth noting that no planetary reflex velocity is identified. This was expected since the amplitudes of the quite distant and low orbital inclination exo-planets would be only fractions of a meter per second which would be masked by the much larger (km\,s$^{-1}$) pulsational velocities and is also beyond the precision attainable by our observations.

\section{Conclusions}
We conclude that the stellar rotational inclination axis has a value i$\gtrsim$40$^\circ$ based on identification of the 1.98\,d$^{-1}$ frequency as an $\ell$=1 $m$=1 mode. This is the strongest pulsation in both photometry \citep{Ze99,Cu06} and spectroscopic radial velocities. Through dynamical analyses it is suspected that the planets are mostly in the same plane with small eccentricities and that the planets orbit inclination axis is $\sim$20$^\circ \pm$10$^\circ$ \citep{Re09,La09}. The current data suggests a misalignment of $\Delta i \gtrsim$20$^\circ$ between the stellar rotational inclination and planetary orbit axes, though more detailed pulsational analyses and better orbital fits are needed before this can be confirmed.

\acknowledgments

We thank A. Moya for helping discussion and useful advices. Wright acknowledges support from the Belgian Federal Science Policy (project MO/33/021). ANC acknowledges support from Comit\'e Mixto ESO-GOBIERNO DE CHILE and from BASAL/ FONDAP project. PDC acknowledges financial support from the Fund for Scientific Research - Flanders (FWO; project G.0332.06). Portions of this work were performed under the auspices of the U.S. Department of Energy by Lawrence Livermore National Laboratory under Contract DE- AC52-07NA27344.

{\it Facilities:} \facility{OHP-1.93m (SOPHIE)}

\clearpage

\end{document}